\documentclass[twocolumn,floatfix,prl,showpacs,superscriptaddress]{revtex4}
\usepackage{graphicx}

\begin{document}
\title{Electron spin resonance in a strong-rung spin-1/2 Heisenberg  ladder. }

\author{A.~N.~Ponomaryov}
\affiliation{Dresden High Magnetic Field Laboratory (HLD-EMFL), Helmholtz-Zentrum Dresden-Rossendorf, D-01328 Dresden, Germany}

\author{M.~Ozerov}
\thanks{Present Address: FELIX Laboratory, Radboud University, 6525 ED Nijmegen, The Netherlands}
\affiliation{Dresden High Magnetic Field Laboratory (HLD-EMFL), Helmholtz-Zentrum Dresden-Rossendorf, D-01328 Dresden, Germany}

\author{L.~Zviagina}
\affiliation{Dresden High Magnetic Field Laboratory (HLD-EMFL), Helmholtz-Zentrum Dresden-Rossendorf, D-01328 Dresden, Germany}

\author{J.~Wosnitza}
\affiliation{Dresden High Magnetic Field Laboratory (HLD-EMFL), Helmholtz-Zentrum Dresden-Rossendorf, D-01328 Dresden, Germany}
\affiliation{Institut f\"ur Festk\"orperphysik, TU Dresden, D-01062 Dresden,Germany}

\author{K.~Yu.~Povarov}
\affiliation{Neutron Scattering and Magnetism, Laboratory for Solid State Physics, ETH Z\"{u}rich, Switzerland}

\author{F.~Xiao}
\thanks{Present Address: Department of Physics, Durham University,  Durham DH1 3LE, United Kingdom}
\affiliation{Neutron Scattering and Magnetism, Laboratory for Solid State Physics, ETH Z\"{u}rich, Switzerland}
\affiliation{Clark University,  Worcester, MA 01610, USA}

\author{A.~Zheludev}
\affiliation{Neutron Scattering and Magnetism, Laboratory for Solid State Physics, ETH Z\"{u}rich, Switzerland}

\author{C.~Landee}
\affiliation{Clark University, 950 Main Street,    Worcester, MA 01610, USA}

\author{E.~\v{C}i\v{z}m\'{a}r}
\affiliation{Institute of Physics, P.J. \v{S}af\'{a}rik University, Ko\v{s}ice, Slovakia}

\author{A.~A.~Zvyagin}
\affiliation{Max Planck Institute for the Physics of Complex Systems, D-01187 Dresden, Germany}
\affiliation{B.~I.~Verkin Institute for Low Temperature Physics and Engineering of the National Academy of Science of Ukraine, Kharkov 61103, Ukraine}

\author{S.~A.~Zvyagin}
\affiliation{Dresden High Magnetic Field Laboratory (HLD-EMFL), Helmholtz-Zentrum Dresden-Rossendorf, D-01328 Dresden, Germany}

\date{\today}

\begin{abstract}

 Cu(C$_8$H$_6$N$_2$)Cl$_2$, a  strong-rung spin-1/2 Heisenberg ladder compound, is probed by means of electron spin resonance (ESR) spectroscopy in the  field-induced gapless phase above $H_{c1}$.  The temperature dependence of the ESR  linewidth is analyzed in the  quantum field theory framework,  suggesting that the anisotropy of magnetic interactions plays a crucial role, determining the peculiar low-temperature ESR linewidth behavior.
In particular,  it is argued  that the uniform Dzyaloshinskii-Moriya interaction (which  is allowed  on the bonds along the ladder legs)  can be the source of this  behavior in  Cu(C$_8$H$_6$N$_2$)Cl$_2$.

\end{abstract}


\maketitle

Quantum spin ladders exhibit a variety of exotic strongly correlated states continuing  to attract a great deal of attention.  One of the most remarkable discoveries on this family of low-dimensional  spin systems  is  superconductivity, observed in  Sr$_{0.4}$Ca$_{13.6}$Cu$_{24}$O$_{41.84}$ under pressure \cite{SC}.  Recently, spin ladders  have been used to address a number of fundamental questions,   related to field-induced phase transitions in quantum matter. In magnetic fields, spin ladders undergo a transition from a disordered gapped to  gapless phase, where   they  can be mapped onto a system of interacting bosons  with  the density of states easily  controlled by the applied magnetic field.  At  low-enough temperatures, when three-dimensional  interactions become important,  the system  undergoes  a  transition into the magnetically ordered state  (often recalled  as to  the magnon  Bose-Einstein condensation \cite{BEC}),  while  above these temperatures (when one-dimensional interactions become dominant)  the  quantum spin ladders  provide a remarkable realization of  the  Tomonaga-Luttinger liquid (TLL) state \cite{TLL}.

Among other spin-Hamiltonian parameters, the anisotropy  of   spin-spin interactions (hereafter magnetic anisotropy) is known to play a  particularly significant role \cite{Gol,Ner}, being, for instance, of crucial importance for the realization of the magnon  Bose-Einstein condensation  in gapped quantum magnets (where the presence of the uniaxial $\rm U(1)$ symmetry,  corresponding  to the global rotational symmetry of the bosonic field phase, is one of the most  essential conditions). Electron spin resonance (ESR) spectroscopy is traditionally recognized as one of the most sensitive tools to probe the magnetic anisotropy  in magnets. In particular, a theory of the low-temperature ESR   was  developed for uniform spin-1/2 Heisenberg antiferromagnetic (AF) chains and spin-1/2 Heisenberg AF chains perturbed by a staggered magnetization \cite{OA-1,Zv-A,OA-2}.  The theory allowed to explain the effect of the anisotropy on the ESR linewidth and field shift in a number of spin-chain systems (see \cite{OA-2,Zvyagin-CuP,Validov} and the references therein).
A new approach to determine  anisotropy parameters from the low-temperature ESR frequency shift in a spin-1/2 strong-rung ladder  has been recently developed by Furuya et al. \cite{Furuya1}. The theory was applied to (C$_5$H$_{12}$N)$_2$CuBr$_4$ (also known as BCPB or (Hpip)$_2$CuBr$_4$), indicating a good agreement with the experimental data \cite{Cizmar}. It was shown also,  that  the anisotropy strongly affects   the ESR linewidth in spin-1/2 strong-leg  ladders \cite{Glazkov}, in particular in the TLL phase \cite{Furuya2}.

The strong-rung spin-1/2 ladder material,  Cu(C$_8$H$_6$N$_2$)Cl$_2$ [hereafter Cu(Qnx)Cl$_2$, where Qnx indicates quinoxaline, C$_8$H$_6$N$_2$] with $J_{r}/k_B =34.2$~K and $J_{l}/k_B =18.7$~K \cite{Povarov}, where $J_r$ and $J_l$ are the exchange coupling constants along the rungs and legs, respectively. This material is characterized by $H_{c1}\simeq 14.3$ T \cite{Povarov,Brian}, which can be easily reached using a superconducting magnet (the second critical field corresponding to the fully spin polarized phase is $H_{c2}\simeq 52$ T \cite{Disser}).  The relatively low $H_{c1}$ makes this compound a perfect model system for studying  spectral and magnetic properties  of strong-rung spin-1/2 ladders  in the field-induced gapless phase, where the  TLL regime can be realized \cite{rem1}.

\begin{figure}[!h]
\centering
\includegraphics[width=0.5\textwidth]{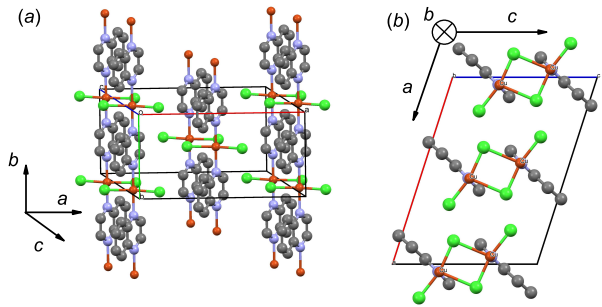}
\caption{\label{fig:Structure}(color online) Two views of the crystal structure of the spin-$1/2$ spin-ladder compound Cu(Qnx)Cl$_2$. The Cu, N, C,  and Cl ions  are shown in red, blue, gray, and green, respectively. The ladders run along the \emph{b}-axis direction.}
\end{figure}

Series of Cu(Qnx)Cl$_2$ single crystals of typically $1$~mm$^3$ size were synthesized at the ETH Z\"{u}rich using slow diffusion in methanol solution, as described in Ref.~\cite{Brian}. The crystal structure of Cu(Qnx)Cl$_2$ is shown schematically  in Fig.~\ref{fig:Structure}. The compound belongs to the monoclinic space group $C2/m$ with unit-cell parameters \emph{a}~=~13.237~\AA, \emph{b}~=~6.935~\AA, \emph{c}~=~9.775~\AA, and $\beta~=~107.88^\circ$ ($Z=4$), measured at room temperature \cite{Somosa, Lindroos}. The spin-1/2 Cu$^{2+}$ ions (shown in red in Fig.~\ref{fig:Structure}) are bridged by  quinoxaline molecules, forming chains along the two-fold rotation axis $b$. Two chains are coupled into a two-leg ladder over two Cl$^-$ ions in each rung.

The samples first were characterized using  the X-band �Bruker Elexsys E500� ESR spectrometer operated at a  frequency of 9.4 GHz. High-quality, twin-free samples were chosen for high-field experiments. At room temperature X-band ESR experiments revealed $g_a = 2.16(1)$, $g_b = 2.03(1)$, and $g_c = 2.09(1)$. High-field  ESR experiments  were performed using a spectrometer, similar to that described in Ref. \cite{Zvyagin_sp}. The spectrometer was  equipped with a 16 T superconducting magnet, transmission-type probe in the Faraday configuration, and VDI radiation sources (product of Virginia Diodes Inc.). The magnetic field was applied along the $b$ axis.

\begin{figure}[!h]
    \centering
    \includegraphics[width=0.5\textwidth]{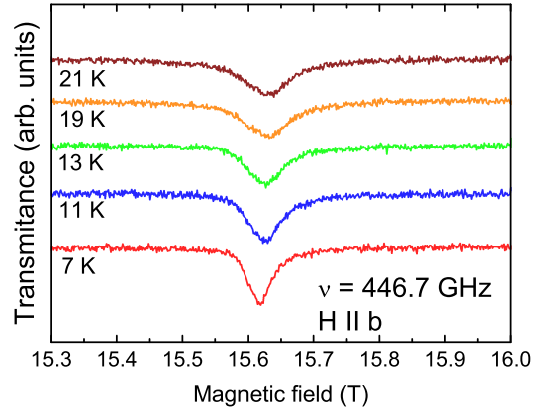}
    \caption{\label{fig:Spectra}(color online) Examples of ESR spectra obtained at 446.7 GHz. }
\end{figure}

\begin{figure}[!h]
    \centering
    \includegraphics[width=0.5\textwidth]{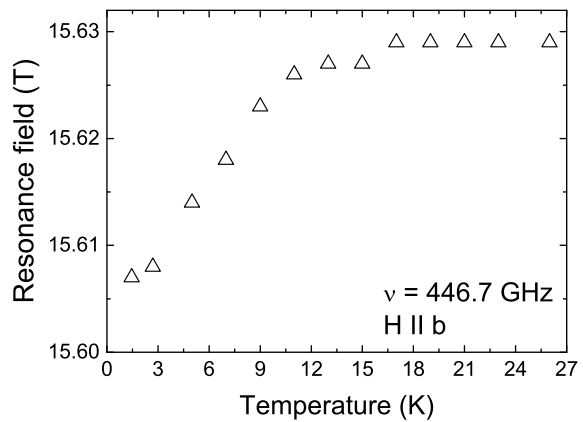}
    \caption{\label{fig:Field} ESR field as function of temperature obtained at 446.7 GHz.}
\end{figure}

\begin{figure}[!h]
    \centering
    \includegraphics[width=0.5\textwidth]{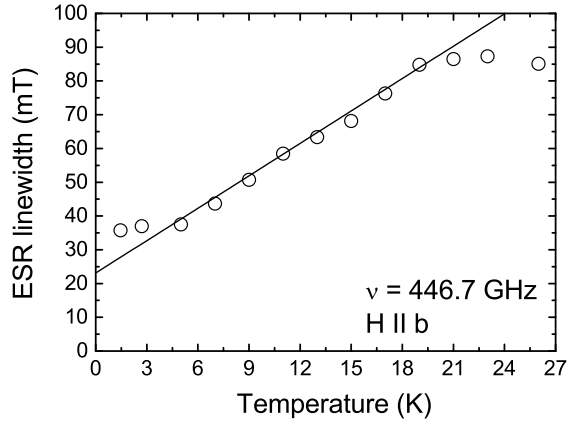}
    \caption{\label{fig:Linewidth} ESR linewidth as function of temperature obtained at 446.7 GHz (symbols). The line is the fit results  using Eq. (11) (see the text for details).}
\end{figure}

Examples of ESR spectra for the frequency 446.7 GHz are shown in Fig.~\ref{fig:Spectra}. The ESR spectra were fit using the Lorentzian lineshape function. With decreasing temperature, starting at $T\sim J_{l}/k_B$ the ESR line shifts towards smaller magnetic fields (Fig.~\ref{fig:Field}), indicating the enhancement of short-range spin correlations. These correlations appear also to be be responsible for the pronounced ESR narrowing revealed  by us in Cu(Qnx)Cl$_2$ (Fig.~\ref{fig:Linewidth}).

As shown in Ref.~\cite{GT}, the Hamiltonian of a spin ladder for $H>H_{c1}$   can be reduced to an effective spin-$1/2$ Heisenberg AF chain Hamiltonian. To set the stage we can describe the low-temperature behavior of the system with the Hamiltonian ${\cal H}$ using the bosonization (or conformal field theory) description, namely
\begin{equation}
    {\cal H} = {\cal H}_0 + {\cal H}_{pert},
\end{equation}
where
\begin{equation}
    {\cal H}_0 = {v\over 2} \int dx [\Pi^2 +(\partial_x \phi)^2]
\end{equation}
is the Hamiltonian of the free boson field $\phi$ and its conjugated momentum $\Pi$. The velocity of low-lying spin excitations, $v$, can be written as
\begin{equation}
    v = {\pi J_{eff} \sqrt{1-\Delta^2}\over 2\arccos \Delta}.
\end{equation}
Here, $J_{eff}$ is the effective exchange interaction in the chain (proportional to the exchange coupling $J_{l}$, \cite{GT}) and $\Delta$ is a parameter describing the uniaxial  anisotropy.

The perturbation can be written as
\begin{equation}
    {\cal H}_{pert} = \lambda \int dx\cos (\sqrt{8\pi K} \phi) \ ,
\end{equation}
with the conformal field theory exponent, $K$, given by
\begin{equation}
    K = {\pi \over \pi - \arccos \Delta}.
\end{equation}

The ESR intensity, $I(\omega,H,T)$, in the standard geometry can be expressed in the framework of the bosonization approach \cite{OA-1} for $H < J_{l}$ (we use the units in which $\hbar = g\mu_B = k_B = 1$) as
\begin{equation}
    I(\omega,H,T) \propto -\chi(T,H)\omega {\mathrm Im}{H^2 +\omega^2\over \omega^2 -H^2 - \Pi(\omega,H,T)} \ ,
\end{equation}
where $\chi(T,H)$ is the static susceptibility of the chain, $\omega$ is the frequency of the $ac$ field and $vq = H$ (where $q$ is the wave vector). In the above expression the self energy $\Pi(\omega,H,T)$ can be calculated in the framework of the perturbation theory with respect to $\lambda$. One can show that
\begin{equation}
    \Pi(\omega,q) = 4\pi K v^2 \lambda^2 [F^r(\omega,q) -F^r(0,0)] \ ,
\end{equation}
where $F^r(\omega,q)$ is the retarded propagator of the boson field given by
\begin{eqnarray}
    &&F^r(\omega,q) \propto -{2^{4K-2}v\over \pi^2T^2}
    \left({\pi T\over v}\right)^{4K} \Gamma^2(1-2K) \times
    \nonumber \\
    \times \nonumber &&{\Gamma(K-{i(\omega +vq)\over 2\pi})\Gamma(K-{i(\omega -vq)\over 2\pi})\over
    \Gamma(1-K-{i(\omega +vq)\over 2\pi})\Gamma(1-K-{i(\omega -vq)\over 2\pi})} \ ,
\end{eqnarray}
where $\Gamma(x)$ is the Gamma function. In the case of  resonance we have to use $\omega = H$. Then, it is clear that the shift of the resonance position due to spin-spin interactions and anisotropy is given by the real part of the self energy, while the ESR linewidth is related to its imaginary part. In particular, the linewidth, $\Delta B$, can be expressed as
\begin{eqnarray}
    &&\Delta B \approx \lambda^2 v^2{\sqrt{\pi} K\over 2^{2K}}T^{4K-3}\left(
    {2\pi \over v}\right)^{4K-2} \times \nonumber \\
    &&{\Gamma({1\over 2}-K)\Gamma^2(K)\Gamma(1-2K)\over \Gamma(1-K)}.
\end{eqnarray}

We can consider the weak  anisotropy $A=\Delta J_{eff}$ of the system as a perturbation with respect to the isotropic antiferromagnetic case $K=1$. Notice that only the real (not effective) magnetic anisotropy reveals itself in the ESR experiments. Then, according to the above and neglecting corrections due to marginal perturbations, such as logarithmic ones, one finds
\begin{eqnarray}
    \Delta B \approx \ \left({A\over J_{eff}}\right)^2T.
\end{eqnarray}
Such a linear dependence between $\Delta B$ and $T$ should exist up to $T\approx$ $J_{eff}$.  Taking into account logarithmic corrections, one obtains
\begin{equation}
    \Delta B (T) \approx A^2 \ln^2 \left({v \over T}\right){T\over v^2}.
\end{equation}

The linear fit of the ESR linewidth behavior for 5~K $<T<$ 21~K  \cite{rem_2}, using the equation
\begin{equation}
    \Delta B_{tot} = \Delta B_{0} + {\delta (\Delta B)\over \delta T}T,
\end{equation}
gives $\Delta B_{TI}=21.7(5)$ mT and ${\delta (\Delta B)\over \delta T}=3.2(1)$ mT/K, for the temperature-independent and temperature-dependent contributions, respectively (Fig.~\ref{fig:Linewidth}). The first, temperature-independent, contribution  can be of the van~Vleck origin. Accordingly to Eq. (9), the second  term of Eq. (11) gives
\begin{equation}
    A~\approx~0.066J_{eff}.
\end{equation}
Assuming $J_{eff}=J_{l}=18.7$ K, one obtains $A\approx 1.2$ K.

One  source of ESR line broadening is the exchange anisotropy, which can be roughly estimated using the formula $D_{E}\approx J_{eff}(\Delta g/g)^2$  (where $\Delta g$ is the deviation of the $g$ factor from  the free electron value and  assuming $J_{eff}=J_{l}$) \cite{Moriya}, giving $D_{E}/k_B\sim 0.1$ K. On the other hand, it is worthwhile to mention that the uniform Dzyaloshinskii-Moriya (DM) interaction is allowed  on the bonds along the ladder legs in  Cu(Qnx)Cl$_2$ by the symmetry. The rough estimate using the formula  $D_{DM} \approx J_{eff}(\Delta g/g)$ \cite{Moriya} gives for the DM interaction  $D_{DM}/k_B \sim 1.5$ K. This value agrees well with the one obtained using Eq. (12), $A \approx 1.2$ K, suggesting that the uniform DM interaction can be the source of the ESR linewidth dependence observed by us in  Cu(Qnx)Cl$_2$.

In conclusion, we have  presented systematic ESR studies of the spin-1/2 strong-rung Heisenberg ladder compound Cu(C$_8$H$_6$N$_2$)Cl$_2$. Employing
the quantum field theory approach, we argue that the   anisotropy of magnetic interactions plays a crucial role, determining the peculiar ESR linewidth behavior at low temperatures.

This work was  supported by Deutsche Forschungsgemeinschaft (DFG, Germany), APVV-0132-11, and VEGA 1/0145/13. We acknowledge the support of the HLD at HZDR, member of the European Magnetic Field Laboratory (EMFL). A.~A.~Z. acknowledges the support from the Institute for Chemistry of V.~N.~Karazin Kharkov National University. Work at ETHZ was partially supported by the Swiss
National Science Foundation, Division 2.


\begin{thebibliography}{99}
\bibitem{rem1} Unfortunately, no detailed data on the tempertaure-field phase diagram of this material are available.
\bibitem{rem_2} It is important to mention that  below 5 K, the accuracy of the ESR linewidth estimation  drops down due to the contribution of a tiny but not-negligible  amount of defects. Above 21 K thermal fluctuations start playing an essential role, similar to that in spin-1/2 Heisenberg chains \cite{OA-2}.
\end{thebibliography}
\end{document}